\documentclass[a4paper,12pt]{article}

\usepackage{amsmath}
\usepackage{amssymb,epsfig}
\usepackage{amsfonts}
\usepackage{graphicx}
\usepackage{float}
\usepackage{latexsym}    
\newcommand{\qed}{{\hfill $\Box$}}

\newcommand{\de}{\delta}

\newcommand{\be}{\begin{equation}}
\newcommand{\ee}{\end{equation}}
\newcommand{\bea}{\begin{eqnarray}}
\newcommand{\eea}{\end{eqnarray}}
\newcommand{\ba}{\begin{array}}
\newcommand{\ea}{\end{array}}

\newcommand{\resetequ}{\setcounter{equation}{0}}

\begin{document} 

\title{Vanishing of Beta Function\\ of Non Commutative
 $\Phi^4_4$ Theory to all orders\footnote{Work supported by ANR grant NT05-3-43374 ``GenoPhy".}}
\author{Margherita Disertori\footnote{Laboratoire de Math\'ematiques Rapha\"el Salem, CNRS UMR 6085,
Universit\'e de Rouen, 76801 Rouen Cedex. E-mail: Margherita.Disertori@univ-rouen.fr}, Razvan Gurau\footnote{Laboratoire de Physique Th\'eorique, CNRS UMR 8627, Universit\'e Paris-Sud XI, 91405 Orsay. E-mail: Razvan.Gurau@th.u-psud.fr}\\
Jacques Magnen\footnote{Centre de Physique Th\'eorique, CNRS UMR 7644,
Ecole Polytechnique F-91128 Palaiseau Cedex, France. E-mail: magnen@cpht.polytechnique.fr}
and Vincent Rivasseau\footnote{Laboratoire de Physique Th\'eorique, CNRS UMR 8627,
Universit\'e Paris-Sud XI, 91405 Orsay. E-mail: Vincent.Rivasseau@th.u-psud.fr}}

\maketitle
 
\begin{abstract}
The simplest non commutative renormalizable field theory, the $\phi_4$ 
model on four dimensional Moyal space with harmonic potential 
is asymptotically safe up to three loops, as shown by 
H. Grosse and R. Wulkenhaar, M. Disertori and V. Rivasseau. 
We extend this result to all orders.
\end{abstract}

\section{Introduction} 

Non commutative (NC) quantum field theory (QFT) may be important for physics 
beyond the standard model and for understanding the quantum
Hall effect \cite{DN}.
It also occurs naturally as an effective regime of string theory \cite{CDS} \cite{SW}.

The simplest NC field theory is the $\phi_4^4$ 
model on the Moyal space. Its perturbative renormalizability 
at all orders has been proved by 
Grosse, Wulkenhaar and followers \cite{GW1}\cite{GW2}\cite{RVW}\cite{GMRV}. 
Grosse and Wulkenhaar solved the difficult problem of ultraviolet/infrared
mixing by introducing a new harmonic potential term 
inspired by the Langmann-Szabo (LS)
duality \cite{LS} between positions and momenta. 

Other renormalizable models of the same kind, including the orientable
Fermionic Gross-Neveu model \cite{V}, have been recently also shown renormalizable at all orders 
and techniques such as the parametric representation have
been extended to NCQFT \cite{GR}.
It is now tempting to conjecture that
commutative renormalizable theories in general have NC renormalizable
extensions to Moyal spaces which imply new parameters. However
the most interesting case, namely the one of gauge theories, still remains elusive.

Once perturbative renormalization is understood, the next problem is to
compute the renormalization group (RG) flow.
It is well known that the ordinary commutative $\phi_4^4$
model is not asymptotically free in the ultraviolet regime. 
This problem, called the Landau ghost or triviality problem affects also quantum electrodynamics.
It almost killed quantum field theory, which was resurrected by the discovery of ultraviolet 
asymptotic freedom in non-Abelian gauge theory \cite{thooft}.

An amazing discovery was made in \cite{GWbeta}:
the non commutative $\phi_4^4$ model does not exhibit any Landau ghost 
at one loop. It is not asymptotically free either. 
For any renormalized Grosse-Wulkenhaar harmonic potential parameter $\Omega_{ren} >0$, 
the running $\Omega$ tends to the special LS dual point $\Omega_{bare} =1$ in the ultraviolet. As a result
the RG flow of the coupling constant is simply bounded \footnote{The Landau ghost can be recovered
in the limit $\Omega_{ren}\to 0$.}. This result was extended up to three loops in \cite{DR}.

In this paper we compute the flow at the special LS dual point $\Omega =1$, and check that 
the beta function vanishes at all orders using a kind of Ward identity inspired by 
those of the Thirring or Luttinger models \cite{MdC,BM1,BM2}.
Note however that in contrast with these models, the model we treat has 
quadratic (mass) divergences. 

The non perturbative construction of the model should combine
this result and a non-perturbative multiscale analysis \cite{GJ}\cite{R}.
Also we think the Ward identities discovered here might be important for the
future study of more singular models such as Chern-Simons or Yang Mills theories,
and in particular for those which have been advocated in connection with 
the Quantum Hall effect \cite{Suss,Poly,HellRaams}.

In this letter we give the complete argument of the vanishing 
of the beta function at all orders in the renormalized coupling, but we assume
knowledge of renormalization and effective expansions as described e.g. in \cite{R},
and of the basic papers for renormalization of NC $\phi^4_4$ 
in the matrix base \cite{GW1,GW2,RVW}.

\section{Notations and Main Result}
\resetequ

We adopt simpler notations than those of \cite{GWbeta}\cite{DR}, and normalize so that $\theta =1$,
hence have no factor of $\pi$ or $\theta$.

The propagator in the matrix base at $\Omega=1$ is
\be \label{propafixed}
C_{mn;kl} = G_{mn} \delta_{ml}\delta_{nk} \ ; \ 
G_{mn}= \frac{1}{A+m+n}\  ,
\ee
where $A= 2+ \mu^2 /4$, $m,n\in \mathbb{N}^2$ ($\mu$ being the mass)
and we used the notations
\be
\de_{ml} = \de_{m_1l_1} \de_{m_2l_2}\ , \qquad m+m = m_1 + m_2 + n_1 + n_2 \ .
\ee

There are two version of this theory, the real and complex one. We focus on the complex case, the result
for the real case folows easily \cite{DR}.

The generating functional is:
\bea
&&Z(\eta,\bar{\eta})=\int d\phi d\bar{\phi}~e^{-S(\bar{\phi},\phi)+F(\bar{\eta},\eta,;\bar{\phi},\phi)}\nonumber\\
&&F(\bar{\eta},\eta;\bar{\phi},\phi)=  \bar{\phi}\eta+\bar{\eta}\phi \nonumber\\
&&S(\bar{\phi},\phi)=\bar{\phi}X\phi+\phi X\bar\phi+A\bar{\phi}\phi+
\frac{\lambda}{2}\phi\bar{\phi}\phi\bar{\phi}
\eea
where traces are implicit and the matrix $X_{m n}$ stands for $m\delta_{m n}$. $S$ is the action and $F$ the external sources. 

We denote $\Gamma^4(0,0,0,0)$ the amputated one particle irreducible four point function  
and $\Sigma(0,0)$ the amputated one particle irreducible two point function 
with external indices set to zero. The wave function renormalization is
$\partial_L \Sigma = \partial_R \Sigma = \Sigma (1,0) - \Sigma (0,0)$ \cite{DR}.
Our main result is:

\medskip
\noindent{\bf Theorem}
\medskip
The equation:
\bea\label{beautiful}
\Gamma^{4}(0,0,0,0)=\lambda (1-\partial_{L}\Sigma(0,0))^2
\eea
holds up to irrelevant terms to {\bf all} orders of perturbation, either as a bare equation with fixed ultraviolet cutoff,
or as an equation for the renormalized theory. In the latter case $\lambda $ should still be understood 
as the bare constant, but reexpressed as a series in powers of $\lambda_{ren}$.

\section{Ward Identities}

Let $U=e^{\imath A}$ with $A$ small. We consider the ``right" change of variables:
\bea
\phi^U=\phi U;\bar{\phi}^U=U^{\dagger}\bar{\phi} \ .
\eea
There is a similar ``left" change of variables. The variation of the action is, at first order:
\bea
\delta S&=&\phi U X U^{\dagger}\bar{\phi}-\phi X \bar{\phi}\approx
\imath\big{(}\phi A X\bar{\phi}-\phi X A \bar{\phi}\big{)}\nonumber\\
&=&\imath A\big{(}X\bar{\phi}\phi-\bar{\phi}\phi X \big{)}
\eea
and the variation of the external sources is:
\bea
\delta F&=&U^{\dagger}\bar{\phi}\eta-\bar{\phi}\eta+\bar{\eta}\phi U-\bar{\eta}\phi 
        \approx-\imath A \bar{\phi}\eta+\imath\bar{\eta}\phi A\nonumber\\
	&=&\imath A\big{(}-\bar{\phi}\eta+\bar{\eta}\phi{)}
\eea
We obviously have:
\bea
&&\frac{\delta \ln Z}{\delta A_{b a}}=0=\frac{1}{Z(\bar{\eta},\eta)}\int d\bar{\phi} d\phi
   \big{(}-\frac{\delta S}{\delta A_{b a}}+\frac{\delta F}{\delta A_{b a}}\big{)}e^{-S+F}\nonumber\\
   &&=\frac{1}{Z(\bar{\eta},\eta)}\int d\bar{\phi} d\phi  ~e^{-S+F}
\big{(}-[X \bar{\phi}\phi-\bar{\phi}\phi X]_{a b}+
       [-\bar{\phi}\eta+\bar{\eta}\phi]_{a b}\big{)} \ .
\eea

We now take $\partial_{\eta}\partial_{\bar{\eta}}|_{\eta=\bar{\eta}=0}$ 
on the above expression. As we have at most two insertions we get only the connected components of the correlation functions.
\bea
0=<\partial_{\eta}\partial_{\bar{\eta}}\big{(}
-[X \bar{\phi}\phi-\bar{\phi}\phi X]_{a b}+
       [-\bar{\phi}\eta+\bar{\eta}\phi]_{a b}\big{)}e^{F(\bar{\eta},\eta)} |_0>_c \ ,
\eea
which gives:
\bea
<\frac{\partial(\bar{\eta}\phi)_{a b}}{\partial \bar{\eta}}\frac{\partial(\bar{\phi}\eta)}{\partial \eta}
-\frac{\partial(\bar{\phi}\eta)_{a b}}{\partial \eta}\frac{\partial (\bar{\eta}\phi)}{\partial \bar{\eta}}
- [X \bar{\phi}\phi-\bar{\phi}\phi X]_{a b}
\frac{\partial(\bar{\eta}\phi)}{\partial \bar{\eta}}\frac{\partial (\bar{\phi}\eta)}{\partial\eta}>_c=0
\ .
\eea
Using the explicit form of $X$ we get:
\bea
(a-b)<[ \bar{\phi}\phi]_{a b}
\frac{\partial(\bar{\eta}\phi)}{\partial \bar{\eta}}\frac{\partial (\bar{\phi}\eta)}{\partial\eta}>_c=
<\frac{\partial(\bar{\eta}\phi)_{a b}}{\partial \bar{\eta}}\frac{\partial(\bar{\phi}\eta)}{\partial \eta}>_c
-<\frac{\partial(\bar{\phi}\eta)_{a b}}{\partial \eta}\frac{\partial (\bar{\eta}\phi)}{\partial \bar{\eta}}> \ ,
\eea
and for $\bar{\eta}_{ \beta \alpha} \eta_{ \nu \mu}$ we get:
\bea
(a-b)<[ \bar{\phi}\phi]_{a b} \phi_{\alpha \beta} 
\bar{\phi}_{\mu \nu }>_c=
<\delta_{a \beta}\phi_{\alpha b} \bar{\phi}_{\mu \nu}>_c
-<\delta _{b \mu }\bar{\phi}_{a \nu} \phi_{\alpha \beta}>_c
\eea

We now restrict to terms in the above expressions which are planar with a single external face,
as all others are irrelevant. Such terms have $\alpha=\nu$, $a=\beta$ and $b=\mu$. 
The Ward identity reads:
\bea\label{ward2point}
(a-b)<[ \bar{\phi}\phi]_{a b} \phi_{\nu a} 
\bar{\phi}_{b \nu }>_c=
<\phi_{\nu b} \bar{\phi}_{b \nu}>_c
-<\bar{\phi}_{a \nu} \phi_{\nu a}>_c
\eea
(repeated indices are not summed). There is a similar Ward identity obtained with the left transformation
and a $\phi\bar{\phi}$ insertion.

\begin{figure}[hbt]
\centerline{\epsfig{figure=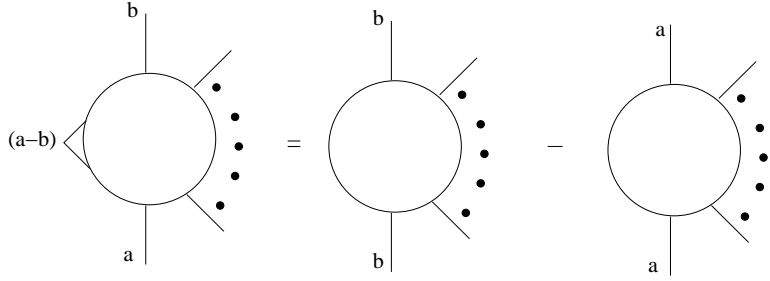,width=10cm}}
\caption{The Ward identity}\label{fig:Ward}
\end{figure}

Deriving once more we get:
\bea
&&(a-b)<[\bar{\phi}\phi]_{a b}\partial_{\bar{\eta}_1}(\bar{\eta}\phi)
\partial_{\eta_1}(\bar{\phi}\eta) \partial_{\bar{\eta}_2}(\bar{\eta}\phi)
\partial_{\eta_2}(\bar{\phi}\eta) >_c=\\
&&<\partial_{\bar{\eta}_1}(\bar{\eta}\phi)
\partial_{\eta_1}(\bar{\phi}\eta)\big{[}
 \partial_{\bar{\eta_2}}
 (\bar{\eta}\phi)_{ab}\partial_{\eta_2}(\bar{\phi}\eta)-\partial_{\eta_2}(\bar{\phi}\eta)_{a b}
 \partial_{\bar{\eta}_2}(\bar{\eta}\phi) \big{]}>_c+1 \leftrightarrow 2 \ .\nonumber
\eea
Take $\bar{\eta}_{1~\beta \alpha}$, $\eta_{1~ \nu\mu}$, $\bar{\eta}_{2~\delta \gamma}$ and $\eta_{2~\sigma \rho}$.
We get:
\bea
&&(a-b)<[\bar{\phi}\phi]_{ab}\phi_{\alpha \beta}\bar{\phi}_{\mu \nu}\phi_{\gamma \delta}
\bar{\phi}_{\rho \sigma}>_c\\
&&=<\phi_{\alpha \beta}\bar{\phi}_{\mu \nu}  \delta_{a \delta}\phi_{\gamma b}\bar{\phi}_{\rho \sigma}>_c
-<\phi_{\alpha \beta}\bar{\phi}_{\mu \nu}\phi_{\gamma \delta}\bar{\phi}_{a \sigma}\delta_{b \rho}>_c+
\nonumber\\
&&<\phi_{\gamma \delta}\bar{\phi}_{\rho \sigma}  \delta_{a \beta}\phi_{\alpha b}\bar{\phi}_{\mu \nu}>_c
-<\phi_{\gamma \delta}\bar{\phi}_{\rho \sigma}\phi_{\alpha \beta}\bar{\phi}_{a \nu}\delta_{b \mu}>_c \ .
\nonumber
\eea
Again neglecting all terms which are not planar with a single external face leads to
\bea\label{ward4point}
(a-b) <\phi_{\alpha a}[\bar{\phi}\phi]_{ab}\bar{\phi}_{b\nu}\phi_{\nu \delta}\bar{\phi}_{\delta \alpha}>_c=
<\phi_{\alpha b}\bar{\phi}_{b \nu}\phi_{\nu \delta}\bar{\phi}_{\delta\alpha}>_c-
<\phi_{\alpha a}\bar{\phi}_{a \nu}\phi_{\nu \delta}\bar{\phi}_{\delta\alpha}>_c \ .
\nonumber
\eea
Clearly there are similar identities for $2p$ point functions for any $p$.

\section{Proof of the Theorem}

\begin{figure}[hbt]
\centerline{\epsfig{figure=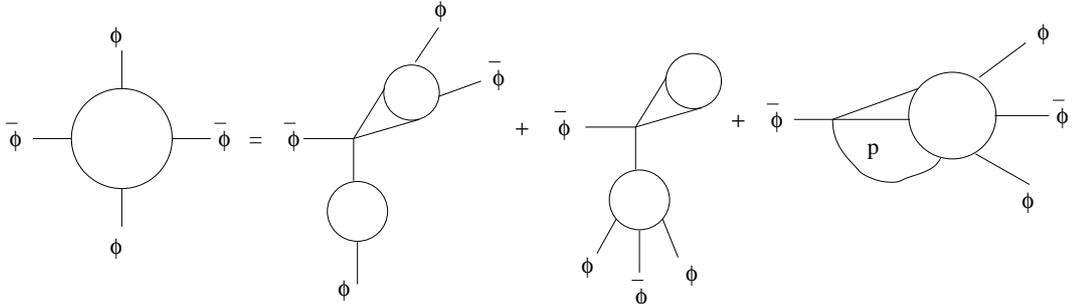,width=14cm}}
\caption{The Dyson equation}\label{fig:dyson}
\end{figure}

We will denote $G^{4}(m,n,k,l)$ the connected four point function restricted to the planar single-border case, where $m$, $n$ ... are the indices of the external borders in the correct cyclic order). $G^{2}(m,n)$ 
is the corresponding connected planar single-border two point function and $G_{ins}(a,b;...)$ 
the planar single-border 
connected functions with one insertion on the left border where the matrix index jumps from $a$ to $b$.
All the identities we use, either Ward identities or the Dyson equation of motion
can be written either for the bare  theory or for the theory with complete mass renormalization, which is the one considered in \cite{DR}. In the first case the parameter $A$ in (\ref{propafixed}) is the bare one, $A_{bare}$
and there is no mass subtraction. In the second case the parameter $A$ in (\ref{propafixed}) 
is $A_{ren}= A_{bare} - \Sigma(0,0)$, and every two point 1PI subgraph is subtracted at 0 external indices\footnote{These mass subtractions need not be rearranged into forests 
since 1PI 2point subgraphs never overlap non trivially.}

Let us prove first the Theorem in the mass-renormalized case, then in the next subsection
in the bare case. Indeed the mass renormalized theory is the one used in \cite{DR}: it is free from any quadratic divergences, and remianing logarithmic subdivergences in the ultra violet  cutoff
can then be removed easily by passing to the ``useful" renormalized effective series, 
as explained in \cite{DR}. 

We analyze a four point connected function $G^4(0,m,0,m)$ with index $m \ne 0$ on the right borders. 
This explicit break of left-right symmetry is adapted to our problem.

Consider a $\bar{\phi}$ external line and the first vertex hooked to it. 
Turning right on the $m$ border at this vertex we meet a new line. If we cut it the graph
may fall into two disconnected components having either 2 and 4 or 4 and 2 external lines
 ($G^{4}_{(1)}$ and $G^{4}_{(2)}$ in Fig. \ref{fig:dyson}) or it may remain connected,
 in which case the new line was part of a loop ($G^{4}_{(3)}$ in Fig. \ref{fig:dyson}). Accordingly
\bea
\label{Dyson}
 G^4(0,m,0,m)=G^4_{(1)}(0,m,0,m)+G^4_{(2)}(0,m,0,m)+G^4_{(3)}(0,m,0,m)\, .
\eea    
The second term  $G^{4}_{(2)}$ is zero after mass renormalization of the two point insertion since
it has a two point subgraph with zero external border. 
We will prove that $G^{4}_{(1)}+G^{4}_{(3)}$ yields 
$\Gamma^4=\lambda (1-\partial \Sigma)^2$ after amputation of the four external propoagators.

Start with $G^{4}_{(1)}$. It is of the form:
\bea
G^4_{(1)}(0,m,0,m)=\lambda C_{0 m} G^{2}(0, m) G^{2}_{ins}(0,0;m)\,.
 \eea

By the Ward identity we have:
\bea
G^{2}_{ins}(0,0;m)&=&\lim_{\zeta\rightarrow 0}G^{2}_{ins}(\zeta ,0;m)=
\lim_{\zeta\rightarrow 0}\frac{G^{2}(0,m)-G^{2}(\zeta,m)}{\zeta}\nonumber\\
&=&-\partial_{L}G^{2}(0,m) \, ,
\eea
and as $\partial_L C^{-1}_{ab}=\partial_R C^{-1}_{ab}=1$ and  $G^{2}(0,m)=[C_{0m}^{-1}-\Sigma(0,m)]^{-1}$ one has:
\bea\label{g41}
G^4_{(1)}(0,m,0,m)&=&\lambda
C_{0m}\frac{C_{0m}C^2_{0m}[1-\partial_{L}\Sigma(0,m)]}{[1-C_{0m}\Sigma(0,m)]
(1-C_{0m}\Sigma(0,m))^2}\nonumber\\
&=&\lambda (C^{D}_{0m})^{4}\frac{C_{0m}}{C^{D}_{0m}}[1-\partial_{L}\Sigma(0,m)]\, .
\eea
The self energy is (again up to irrelevant terms (\cite{GW2}):
\bea
\label{PropDressed}
\Sigma(m,n)=\Sigma(0,0)+(m+n)\partial_{L}\Sigma(0,0) 
\eea 
Therefore up to irrelevant terms:
\bea
C^D_{0m}=\frac{1}{m[1-\partial_{L}\Sigma(0,0)]  + A_{ren}}\, ,
\eea
and
\bea \label{cdressed}
\frac{C_{0m}}{C^D_{0m}}=1-\partial_{L}\Sigma(0,0)+\frac{A_{ren}}{m+A_{ren}}\partial_{L}\Sigma(0,0) \, .
\eea

\begin{figure}[hbt]
\centerline{\epsfig{figure=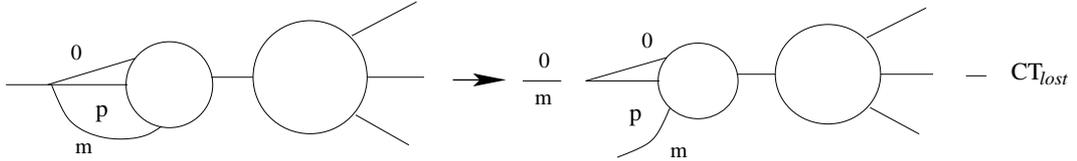,width=14cm}}
\caption{Two point insertion and opening of the loop with index $p$}\label{fig:insertion}
\end{figure}

For the $G^4_{(3)}(0,m,0,m)$ one starts by ``opening" the face which is ``first on the right''
with the $p$ index.  For bare Green functions this reads:
\bea
\label{opening}
G^{4,bare}_{(3)}(0,m,0,m)=C_{0m}\sum_{ p} G^{4,bare}_{ins}(p,0;m,0,m)\, .
\eea
Passing to mass renormalized Green functions one sees that if the face $p$ belonged to a  1PI two point
insertion in $G^{4}_{(3)}$ 
this 2 point insertion disappears on the right hand side of eq.(\ref{opening}) 
(see fig. \ref{fig:insertion})! In the equation for $G^{4}_{(3)}(0,m,0,m)$ one must 
therefore \textit{add its missing counterterm}, so that:
\bea
\label{Open2}
G^4_{(3)}(0,m,0,m)&=& C_{0m}\sum_{p} G^{4}_{ins}(0,p;m,0,m)-CT_{lost}\,.
\eea
The part of the self energy with non trivial right border is called $\Sigma^R$, so that 
the difference $\Sigma - \Sigma^R$ is the generalized left tadpole $\Sigma^L_{tadpole}$.
The missing mass counterterm must have a right $p$ face, so it is restricted to $\Sigma^R$ and is:
\bea\label{lostct}
CT_{lost}=C_{0m}\Sigma^R(0,0)G^{4}(0,m,0,m)\, .
\eea
We compute the value of this counterterm 
again by opening its right face $p$ and using the Ward identity (\ref{ward2point}). We get:
\bea
\label{S2}
\Sigma^R(0,0)&=&\frac{1}{C^D_{00}}\sum_{p}G^2_{ins}(0,p;0)\nonumber\\
    &=&\frac{1}{C^D_{00}}\sum_{p}\frac{1}{p}[G^2(0,0)-G^2(p,0)]\nonumber\\
    &=&\sum_{p}\frac{1}{p} \biggl(1 -\frac{C^{D}_{p0}}{C^D_{00}}\biggr) \, .
\eea
A similar equation can be written also for $\Sigma^R(0,1)$:
\bea
\label{S2new}
\Sigma^R(0,1)&=&\sum_{p}\frac{1}{p} 
\biggl(1 -\frac{C^{D}_{p1}}{C^D_{01}}\biggr) \, .
\eea

We then conclude that:
\bea\label{S3}
CT_{lost}=C_{0m} G^{4}(0,m,0,m) \sum_{p}\frac{1}{p}  
\biggl( 1- \frac{C^{D}_{0p}}{C^D_{00}}  \biggr)  \, .
\eea

But by the Ward identity (\ref{ward4point}): 
\bea
\label{Ward4}
C_{0m} \sum_{p} G^{4}_{ins}(0,p;m,0,m)=C_{0m} \sum_{p} \frac{1}{p}\biggl( G^{4}(0,m,0,m)-G^{4}(p,m,0,m) \biggr)\, ,
\eea
so that subtracting (\ref{S2}) from (\ref{Ward4}) computes:
\bea
\label{G3}
G^4_{(3)}(0,m,0,m) &=&-C_{0m}\sum_{p}\frac{1}{p} \bigl( G^{4}(p,m,0,m) + G^{4}(0,m,0,m)
\frac{C^{D}_{0p}}{C^D_{00}}\bigr) \, .
\eea
The first term in eq (\ref{G3}) is irrelevant, having at least three denominators 
linear in $p$. We rewrite the last term, using  (\ref{S2}), (\ref{S2new})
starting with:
\bea
\partial_{R}\Sigma(0,0)&=&\partial_{R}\Sigma^{R}(0,0)=\Sigma^{R}(0,1)-\Sigma^{R}(0,0)\nonumber\\
&=&\sum_{p}\frac{1}{p} \biggl( \frac{C^{D}_{p0}}{C^D_{00}}
-\frac{C^{D}_{p1}}{C^D_{01}} \biggr) \, ,
\eea
In the second term of the above equation one can change the 
$C^{D}_{p1}$ in $C^{D}_{p0}$ at the price of an irrelevant term. 
Using (\ref{cdressed}) we have:
\bea
\partial_{R}\Sigma(0,0)&=&-[1-\partial_{L}\Sigma(0,0)]\sum_{p}\frac{1}{p}C^{D}_{p0}
\eea
hence
\bea\label{g43}
G^4_{(3)}(0,m,0,m;p)&=&-G^{4}(0,m,0,m)
\frac{A_{ren} \; \partial_{R}\Sigma(0,0)}{(m+A_{ren}) [1-\partial_{L}\Sigma(0,0)]} \ .
\eea
Using (\ref{g41}), (\ref{cdressed}) and (\ref{g43}), equation (\ref{Dyson}) rewrites as:
\bea
\label{final}
&&G^4(0,m,0,m)\Big{(}1+
\frac{A_{ren}\; \partial_{R}\Sigma(0,0)}{(m+A_{ren}) \; [ 1-\partial_{L}\Sigma(0,0)] }\Big{)}
\\
&&=\lambda_{bare} (C^{D}_{0m})^{4}\Big{(}1-\partial_{L}\Sigma(0,0)+\frac{A_{ren}}{m+A_{ren}}\partial_{L}\Sigma(0,0)\Big{)}
[1-\partial_{L}\Sigma(0,m)]\, .\nonumber
\eea
Multiplying (\ref{final}) by $[1-\partial_{L}\Sigma(0,0)]$ and amputating 
four times proves (\ref{beautiful}), hence the theorem.
\qed 

\subsection{Bare identity}

Let us explain now why the main theorem is also true as an identity between bare functions, without
any renormalization, but with ultraviolet cutoff.

Using the same Ward identities, all the equations go through with only few differences:

- we should no longer add the lost mass counterterm in (\ref{lostct})

- the term $G^{4}_{(2)}$ is no longer zero.

- equation (\ref{cdressed}) and all propagators now involve the bare $A$ parameter.

But these effects compensate. Indeed the bare  $G^{4}_{(2)}$ term is the left generalized
tadpole $\Sigma - \Sigma^R$, hence
\begin{equation} \label{newleft}
G^{4}_{(2)}  (0,m,0,m) = C_{0,m} \bigl(  \Sigma(0,m) - \Sigma^R (0,m) \bigr) G^4(0,m,0,m)\; .
\end{equation}
Equation (\ref{cdressed}) becomes up to irrelevant terms
\bea \label{cdressedbare}
\frac{C^{bare}_{0m}}{C^{D,bare}_{0m}}=1-\partial_{L}\Sigma(0,0)+
\frac{A_{bare}}{m+A_{bare}}\partial_{L}\Sigma(0,0) 
- \frac{1}{m+A_{bare}}\Sigma(0,0) 
\eea
The first  term proportional to $ \Sigma(0,m) $ in (\ref{newleft})  combines with 
the new term in (\ref{cdressedbare}), and the second term proportional to $ \Sigma^R(0,m) $ in (\ref{newleft})
is exactly the former ``lost counterterm" (\ref{lostct}). This proves (\ref{beautiful}) in the bare case.

\section{Conclusion}

Since the main result of this paper is proved up to irrelevant 
terms which converge at least like a power of the infrared cutoff, as this infrared cutoff is lifted towards infinity,
we not only get that the beta function vanishes in the ultraviolet regime, but that it 
vanishes fast enough so that the total flow of the coupling constant is bounded. 
The reader might worry whether this conclusion is still true for the full model which has 
$\Omega_{ren} \ne 1$, hence no exact conservation of matrix indices along faces.
The answer is yes, because the flow of $\Omega $ towards its ultra-violet limit
$\Omega_{bare}=1$ is very fast (see e.g. \cite{DR}, Sect II.2). 

The vanishing of the beta function is a step towards a full non perturbative construction 
of this model without any cutoff, just like e.g. the one of the Luttinger model \cite{BGPS,BM1}.
But NC $\phi^4_4$ would be the first such \textit{four dimensional} model, and the only one
with non logarithmic divergences. Tantalizingly, quantum field theory might actually behave 
better and more interestingly on non-commutative than on commutative spaces.

\subsubsection*{Acknowledgment}
We thank Vieri Mastropietro for very useful discussions.

\end{document}